# Some peculiarities of equilibrium between springs and charges

*Michael B. Partensky*,

Department of Chemistry, Brandeis University, Waltham, MA 02453

**partensky@rcn.com**

Common sense tells us that a small variation in external conditions results in a correspondingly small physical response. For example, a tiny increase in the force applied to a spring causes a bit of additional stretching, while moving uphill a few meters results in only a miniscule change in barometric reading. Such behavior, when equilibrium properties response smoothly to variations of external forces or parameters, can be considered "normal".

However, in real life things quite often don't work that way. For example, increasing the load on a branch can result in its suddenly breaking with a painful fall. Gradual decrease in the temperature causes an abrupt transition of water from liquid to solid. When crossing Cape Cod, even at its narrowest (about a mile), many are surprised by the dramatic change in ocean temperature (from comfortably warm to freezing cold).

The explanation of such abnormalities often requires very complex physics and mathematics. Some aspects of these, including the "butterfly effect", are well publicized, thanks to Dr. Malcolm, a charismatic scientist from the "Jurassic Park". However, examples of catastrophic changes in response to a gradual variation of external conditions can also be discussed with no more than simple high school physics. An illustration can be found in the challenging problems offered to our old acquaintances [1], Jeff and Fred, by Mrs. Walker, their AP physics teacher. ( See also [2] for another, slightly more complex, example that still requires only High School physics).                    .

## The statement of the problem

There are two charges, $Q$ and $q$. Both the position and value of $Q > 0$ are fixed while $q$ has a varying value and is attached to an insolating spring with a spring constant $\Gamma$. If $q = 0$, the equilibrium distance $l$ between the charges is equal to $l_0$ (see Fig. 1). For any $z > 0$, find the value of $q$ for which $l = zl_0$. Verify the stability of equilibrium corresponding to the $q$ that you found and analyze the solution.

<u>Challenge</u>  Assume that charges $Q$ and $q$ are surrounded by the hard shells of finite radiuses, restricting the distance of their closest approach to a certain value $z_{min}$   (1) How the equilibrium properties of this electromechanical system depend on $z_{min}$? (2) How the behavior of this system can be affected by the uniform electrostatic field applied along the line connecting the charges.

## The boys discuss the problem

First, Fred and Jeff realized that in equilibrium the electric and elastic forces should be in balance. The electric (Coulomb) force acting on $q$ is $F_Q = k\dfrac{Qq}{l^2}$ where $k$ is the electrostatic constant. The positive direction of $F_Q$, corresponding to $q > 0$ (repulsion, $z \geq 1$), is the direction from $Q$ to $q$. The elastic force defined by Hook's law is $F_H = -\Gamma(l - l_0)$

The condition that the resultant force becomes zero leads to the equation

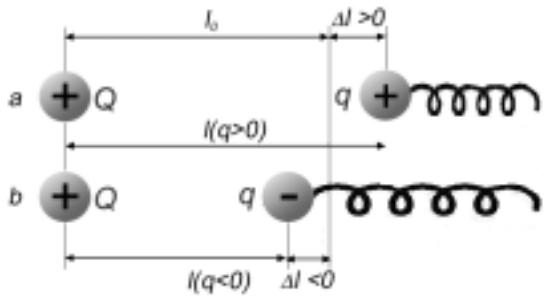

$$k\dfrac{Qq}{l^2} - \Gamma(l - l_0) = 0 \qquad (1)$$

**Fig. 1: The "experimental setting" with similar (a) and opposite (b) charges**

Introducing the dimensionless distance $z = l/l_0$ and considering $l > 0$ they decided to rewrite this equation as

$$\gamma q = z^2 (z - 1) \qquad (z > 0) \qquad (2)$$

where $\gamma = \dfrac{kQ}{l_0^3 \Gamma}$. Then Jeff turned on his graphical calculator and punched the keys.

The resulting plot looked quite surprising (see Fig. 2).

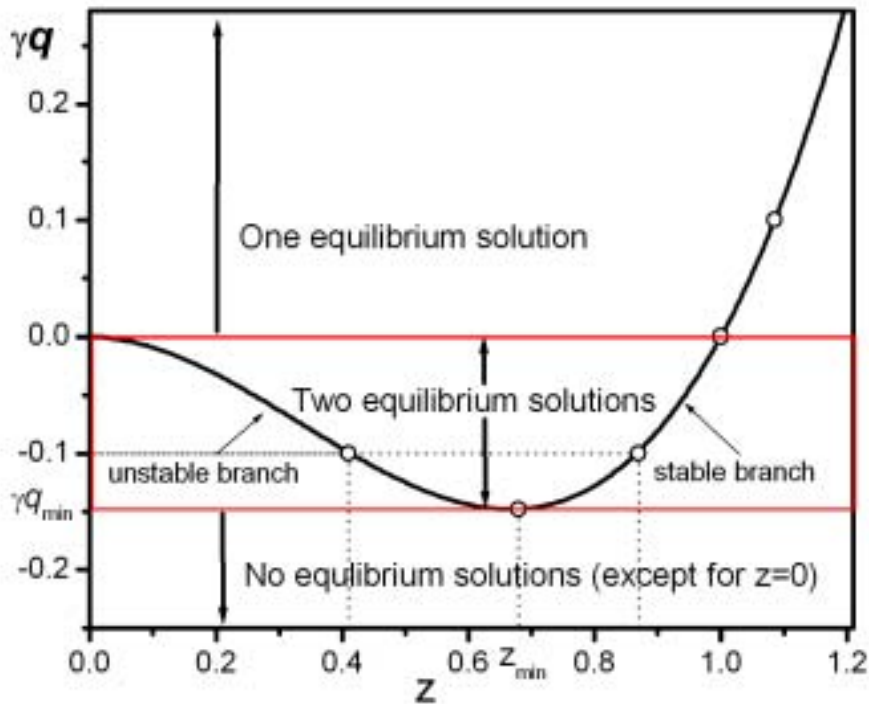

Fig. 2 q(z) dependencies defined from Eq. 2.

Although some of its features, such as behavior at $z \geq z_{min}$, looked very reasonable (you can ask your students to explain a continuous growth of $q(z)$ in this range ), some others seemed to be really unusual. Thus, the boys expected that, when the negative charge $q$ grew more negative, it would continuously approach $Q$ until the charges come in direct contact ($z = 0$). In other words, they expected the curve $q(z)$ to continuously decay until $z = 0$ was reached at some $q < 0$. Apparently, this was not the case. The allowed values of $q$ appeared to be constrained from below by

$q_{min} \sim -0.15/\gamma$ which did not correspond to $z = 0$. To the contrary, the corresponding separation $z_{min}$ was quite big, $\sim 0.7$ [3]. In addition, for each value $q_{min} < q < 0$, there were two (not one) equilibrium roots: $z' > z_{min}$, and $z'' < z_{min}$. The boys had to decide which of two solutions was the "real one".

Confronted with these puzzles, Jeff suddenly remembered another problem that they had solved in class.

That time they were trying to find the equilibrium position of a negative charge in the presence of two equal positive charges. They suggested at first that the negative charge will reside right in the midpoint, but Mrs. Walker asked them to check if the equilibrium was stable. For this purpose, they drew the position dependency of combined electric force acting on the charge. Although the forces were balanced in the midpoint, even a tiny displacement in either direction resulted in a force pushing the charge away from the equilibrium point. It was like a knife balancing on its tip. The only stable configuration corresponded to the negative charge resting right on one of two positive charges. This memory produced the idea of analyzing the force as a function of $z$, which turned out to be fruitful. They rewrote the total force $F = k\frac{Qq}{l^2} - l + l_0$

containing electric and elastic components, as

$$f = \gamma \frac{q}{z^2} - z + 1 \qquad (3)$$

(here $f = F/\Gamma l_0$ is dimensionless force).

In seconds they drew the profiles $f(z)$ shown in Fig. 3. Now the whole picture became almost transparent! The points $z_i$ ($f(z_i) = 0$) are the roots of Eq. 2 for the corresponding values of $q$. If the slope of $f(z)$ at $z = z_i$ is negative, then the equilibrium is stable (the force resists the disturbance)[4]. Such are the solutions $z_1$, $z_2$ and $z_3'$. On the other hand, the positive slope indicates that the equilibrium is unstable. The unstable (fictitious) equilibrium solutions $z''$ exist for any negative $q > q_{min}$. While the elastic force grows very smoothly (~ $1-z$) when $z$ decreases, the electrostatic attraction skyrockets (~$1/z^2$) at small $z$. Consequently, for any $q < 0$ a domain of separations $0 < z < z''$ exists where the attraction dominates over the elastic force, and the charges collapse. The unstable equilibrium solution $z''$ is the upper boundary of such a domain. This domain of instability becomes wider as $q$ grows more negative. Finally, at $q \leq q_{min}$ the electrostatic attraction prevails at all separations, and the only equilibrium location for $q$ becomes right on the top of $Q$.

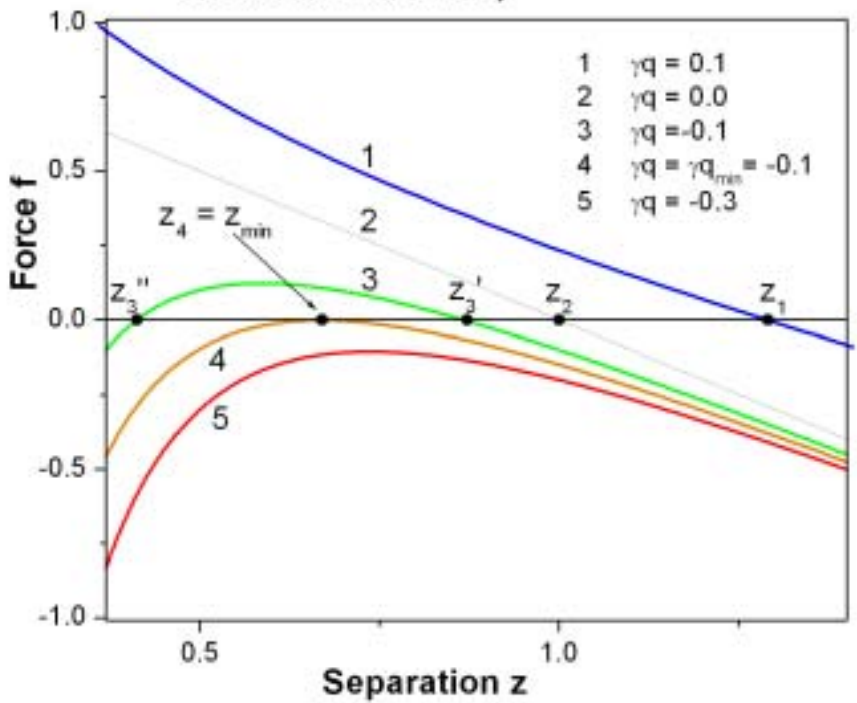

Fig. 3 The force profiles f(z) defined from Eq. 3 for various values of q

1  γq = 0.1
2  γq = 0.0
3  γq = -0.1
4  γq = γq$_{min}$ = -0.1
5  γq = -0.3

"That's amazing!" – Fred was excited- "The equilibrium completely disappears far before the charges touch each other. I still have to digest this idea. This problem is really fantastic!

"It was a good one", responded Jeff. "I wonder if the same thing happens with two magnets. You can't bring magnets close to each other without their jumping together. I got a cool magnetic construction set for my birthday, and we can do all kinds of experiments. But for now let's go out and play ball. We can write it all down after the game. And tomorrow we can deal with the challenge".

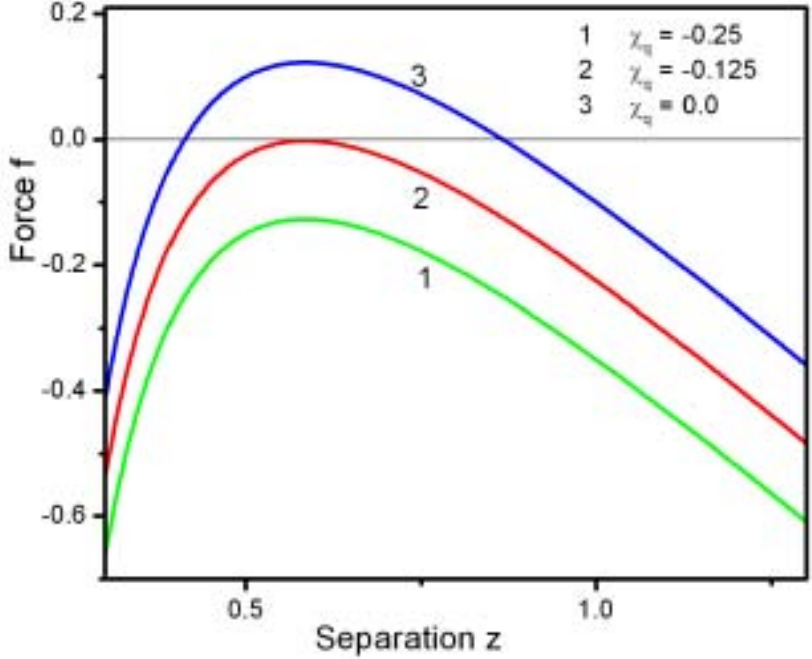

Fig. 4 The force profiles f(z) defined from Eq. 3 for $\gamma q$ =0.1 and three values of the electric field

1  $\chi_q$ = -0.25
2  $\chi_q$ = -0.125
3  $\chi_q$ = 0.0

## The challenge

Let us now outline the solution to the challenge, focusing on the effect of the electric field. According to (1), z is restricted from below by the "distance of the closest approach", $z_{min}$. The "hard shells" prevent Q and q from collapse. If joined, they can be now pulled apart by a moderate external electric field (unlike the point charges considered above that can not be separated after they collapse). This brings our model closer to the molecular "electro-elastic" systems, where different (finite-size) charged groups are attached to the molecular "springs". The changes in the applied voltage result in changes in the molecular conformation. Such devices can be important, for instance, in membrane electroporation and in the voltage gating of ion channels [5,6]. In presence of the electric field $F$, the Eq. 3 should be rewritten as

$$f = \gamma \frac{q}{z^2} - z + 1 + \chi_q \qquad (4)$$

where $\chi_q = \frac{qF}{\Gamma l_0}$ is the dimensionless electric force. The effect of the electric field becomes especially important for $\gamma q < 0$. To be specific, we pick $\gamma q = -0.1$ corresponding to the curve 3 of Fig. 3, and chose $z_{min}$ =0.3 (the profiles $f(z)$ shown in Fig. 4 are restricted to $z \geq 0.3$). Fig 4 demonstrates the important "switching" effect of the electric field: a strong and abrupt change in the geometrical parameters (the separation z between the charges) in response to a continuous variation of the electric field. Suppose that the equilibrium field across the (polarized) membrane corresponds to $\chi_q = -0.25$ ( curve 1). Depolarizing the membrane does not result in any qualitative changes while $\chi_q \leq \chi_q^{cr} = -0.125$: q resides at z=0.3, the position corresponding to the lowest energy configuration. An additional equilibrium located at z= $z_{eq}$ (such that $f(z_{eq})$=0, $f'(z_{eq})$<0 ) appears when $\chi_q$ exceeds $\chi_q^{cr}$. It is separated by a finite distance from $z_{min}$, and moves further away from it with $\chi_q$ increasing. Assuming that the transition occurs when the energy "wells" become equally deep, one can find that the corresponding $\chi_q \sim 0$ (Fig. 4, curve 3). This transition is accompanied by the abrupt and significant increase in the separation between the charges. As previously mentioned, such a phenomena can be related to membrane instabilities and voltage gating of ion channels. Further analysis of the latter relation can be offered as a research project aiming to introduce students to the physical principles behind the "Life's transistors"[7].

## A few comments

Fred and Jeff were able to solve a problem which is closely related to complex physical behavior, associated with phase transitions, bifurcations and catastrophes. In these phenomena, a smooth variation of an external parameter (such as temperature, pressure, etc) leads to a sudden and dramatic changes in the equilibrium state. The instabilities are related to non-linear properties and their analysis typically requires advanced mathematical tools. Probably, for this reason, such problems are avoided even in Honors and AP Physics Textbooks. For instance , many variations of a problem with two equally charged bodies suspended on two strings are quite popular [8], while I was unable to find a single discussion of this problem with two opposite charges.

To avoid the difficulties of solving a cubic equation which would require dealing with complex numbers, Mrs. Walker used a trick. Instead of asking how the equilibrium distance depends on charge she reversed the question. Her students were challenged to *find a charge corresponding to a given equilibrium distance* that required only simple and direct calculations. Previous experience directed Jeff and Fred to a graphical representation of the problem, and made the whole discussion more focused and enjoyable. The same approach can be applied to various problems, including the above problem with the opposite charges suspended at a finite separation.

A similar problem would result from replacing the charges with magnets. The attraction between magnets can typically lead to a behavior similar to that considered above. Students can be asked how the equilibrium distance depends on the position of the fixed end of the spring. Such a problem allows a simple experimental analysis. For my experiments I used a wonderful construction toy, the "Rogers Connections", which is available in many stores and on the Internet (I bought it in the neighboring "The Construction Site" store http://www.constructiontoys.com/). It consists of plastic struts tipped with small high-power magnets and steel balls. The only addition that I needed was a soft spring. Enjoy!

## Acknowledgement:

Many thanks to John Griffin and Jordan L. Wagner for their valuable comments and suggestions. I am grateful to Vitaly Feldman, Peter C. Jordan and Vladimir Dorman for our discussions of instabilities and phase transitions at electrified interfaces, membranes and ion channels. John Griffin also designed and built a simple device demonstrating the peculiarities of "magneto-elastic" behavior.

*References*

1. Partensky M. B. "Two boys and a can of Coca-Cola", The Physics Teacher, <u>40</u>, 106, 2002
2. Partensky M. B. "The elastic capacitor and its unusual properties" ArXiv: physics/0208048 (2002)
3. At this point, as a simple exercise in calculus, students can be asked to find $z_{min}$ and $\gamma q_{min}$ using the minimum condition $d(\gamma q)/dz = 0$. They will discover that $z_{min} = 2/3 \approx 0.67$ and $\gamma q_{min} = 4/27 \approx 0.148$.
4. The students can be reminded that the slope of *f(z)* is described by the derivative $\frac{df}{dz}$. Given the relation between the force and the energy $E(z)$, $f = -\frac{dE(z)}{dz}$, they will find that the negative slope $\frac{df}{dz} < 0$ corresponds to $\frac{d^2 E(z)}{dz^2} > 0$. Combined with the force balance condition $\frac{dE(z)}{dz} = 0$, this indicates that $E(z)$ has minimum (the equilibrium is stable).
5. Partenskii M. B., Dorman V. and Jordan P.C. "Electro-elastic capacitative coupling is a possible mechanism for voltage gating", Biophys. J. <u>70</u> (2), 203, 1998
6. Partenskii M. B. and Jordan P.C. "Electroelastic instabilities in double layers and membranes", In. "Liquid interfaces in chemical, biological and pharmaceutical applications", Volkov A. G. (Editor), 2001, pp. 51-82.
7. Sigworth, F. J. "Life's transistors", Nature, <u>423</u>, 21, 2003.
8. Giancoli, D.C. "Physics. Principles with applications", 5-th ed. (Prentice Hall, Upper Saddle River, NJ, 1997).